\documentclass[12pt,twoside,a4paper]{posnav}
\usepackage{epsf,exscale,times}
\usepackage[english]{babel}
\usepackage{graphicx}
\usepackage{amsmath}
\usepackage{fancyhdr}

\newcommand {\vsp}   {\vspace*}

\def\title#1{\vsp{-16mm}\begin{center}\Large\bf{#1}\end{center}\vsp{0mm}}
\def\author#1{{\begin{center}\textbf{#1}\end{center}\vspace{-1mm}}}
\def\address#1{\vsp{-3mm}\begin{center}\baselineskip12pt\normalsize{#1}\end{center}\vsp{-1mm}}

\def\abstract#1{{\vspace{-5mm}
    \begin{center}
      \begin{minipage}{0.85\textwidth}
        \noindent\bf \textit{Abstract:}
        \small\rm\emph{#1}
				\vsp{-0.5em}
      \end{minipage}
    \end{center}
}}

\def\authorsheadline=#1{\global\def\@authorsheadline{#1}}
\global\def\@authorsheadline{}

\def\TeX{T\kern-.1667em\lower.5ex\hbox{E}\kern-.125emX}

\def\LaTeXG{{\rm L\kern-.36em\raise.3ex\hbox{\sc a}\kern-.15emT\kern-.1667em\lower.7ex\hbox{E}\kern-.125emX}}
\def\LaTeXK{{\it L\kern-.24em\raise.4ex\hbox{\scriptsize \it A}\kern-.20emT\kern-.1667em\lower.5ex\hbox{E}\kern-.125emX}}

\usepackage{booktabs}		
\usepackage{longtable} 		
\usepackage{multirow}		
\usepackage{threeparttable}

\usepackage{subcaption}

\begin{document}

\fancypagestyle{firststyle}
{
   \fancyhf{}
   \lfoot{ \footnotesize{Positioning and Navigation for Intelligent Transport Systems POSNAV 2024, October 1 - 2, 2024, Weimar\break     		\copyright  2024 DGON} }
   \rfoot{ \footnotesize {\thepage} }
}

\thispagestyle{firststyle}
\fancyhf{}
\renewcommand{\headrulewidth}{0pt}
\renewcommand{\footrulewidth}{1pt}
\renewcommand{\footskip}{50pt}

\pagestyle{fancy}
\fancyfoot[RO,LE]{ \footnotesize {\thepage} }

\title{Near-Range Environmental Perception for Inland Waterway Vessels: A Comparative Study of LiDAR and Automotive FMCW RADAR Sensors}


\author{
Robin Herrmann$^{*}$, Soumalya Bose$^{*}$,
Iulian Filip$^{**}$, Daniel Medina$^{**}$, Ralf Ziebold$^{**}$,
Sebastian Gehrig$^{***}$, Thilo Lenhard$^{***}$,
Markus Gardill$^{*}$}
\address{
    $^{*}$Electronic Systems and Sensors\\BTU Cottbus-Senftenberg\\
    Cottbus, Germany\\
		email: robin.herrmann@b-tu.de\\[2mm]
    $^{**}$Nautical Systems, Institut of Communications and Navigation\\ German Aerospace Center (DLR)\\
    Neustrelitz, Germany\\
		email: iulian.filip@dlr.de\\[2mm]
    $^{***}$InnoSenT GmbH\\
    Donnersdorf, Germany\\
		email: thilo.lenhard@innosent.de
		}

\abstract{
  Advancing towards high automation and autonomous operations is crucial for the future of inland waterway transport (IWT) systems.
  These systems necessitate robust and precise onboard sensory technologies that can perceive the environment under all weather conditions,
  including static features for local positioning techniques such as Simultaneous Localization and Mapping (SLAM). Traditional marine RADAR,
  mandatory on vessels and operating in the 9300-9500 MHz frequency band, can cover ranges from 15 to 1200 meters but are inadequate for detecting closer objects,
  making them unsuitable for automated docking maneuvers, lock entry, or bridge undercrossings.
  This necessitates the development of reliable close-range sensor technology that functions effectively in all weather conditions.\\
  In present research works on vessel automation, LiDAR sensors, operating in the near-infrared range,
  are used predominantly to detect the immediate surroundings of vessels but suffer significant degradation in poor visibility.
  Conversely, automotive RADAR sensors, utilizing the 76-81 GHz frequency band, can detect objects from a few centimeters to up to 200 meters, even in adverse conditions.
  These sensors are commonly used in advanced autonomous road traffic systems and are evaluated in this study for their suitability in inland navigation and maneuvering.\\  
  This paper discusses a distributed sensor network of four compact automotive frequency-modulated continuous-wave (FMCW) radars mounted on a cabin boat as a test platform.
  Initial field experiments demonstrate the RADAR network's ability to perceive close-range static environmental features around the boat in inland waters.
  The paper also provides a comparative analysis of the environmental detection capabilities of automotive RADAR and LiDAR sensors.
}


\section{Introduction}
Inland waterways have great potential to be used as transportation routes for heavy and bulky goods, 
which can relieve congestion on the roads in and around cities.  
In order to use them efficiently, progress towards a high degree of automation and autonomous operation would be crucial for the future of these intelligent transportation systems (ITS).
But waterways are often narrow and many bridge undercrossings and ship locks make autonomous operation of ships difficult.
Therefore, precise sensors on board are required to perceive a vessel's surroundings.



Usually LiDAR sensors are used to obtain very detailed detections of objects in the close range of a vessel.
Despite the azimuthal $360^{\circ}$ coverage, several such sensors are often installed around a vessel due to the boat's dimensions or 
because of its narrow field of view in the vertical direction.

These lidars work with light beams in the near-infrared range and therefore 
therefore have their limitations in poor visibility conditions such as heavy rain, fog or snow.
To overcome the limited visibility, scanning marine radars are generally used, which operate with electromagnetic waveforms in the frequency band from 9.3 to 9.5 gigahertz.
They have a fairly long range of up to 1 km, but due to their pulsed operating principle 
their high position and narrow field of view in the direction of elevation, they are not able to detect objects at close range.

As similar challenges need to be solved for autonomous driving in the automotive sector, our project 
intelligent industrial radar sensor networks for the Spree-Oder waterways (iiRadarSOW) 
is investigating the use of automotive radars as a way of supporting the perception of a vessel's surroundings.
The aim is to perceive the surroundings of a boat in a 360-degree close range to support autonomous navigation and maneuvering.
The commonly used LiDAR sensors are used as a reference.

Nowadays, these vehicle radars are mainly used for advanced driver assistance systems (ADAS) and 
at least four RADAR sensors are mounted on the sides of the vehicle to cover a 360-degree area.
The range of the radar sensors extends from a few centimetres to help with parking, for example, up to 200 meters to detect situations on the freeway.
A high bandwidth is required to detect objects with a resolution in the centimeter range. They therefore operate in the frequency band from 76 to 81 Ghz.

Automotive RADAR sensors can work in difficult weather conditions, just like the scanning marine RADAR.
However, automotive RADAR sensors must be compact enough to be installed in a car, unlike a scanning RADAR like the marine RADAR.
This inevitably compromises the antenna aperture and, as a result, the angular resolution.
The small aperture also results in a broad beamwidth, which means that automotive RADARs do not operate in a scanning mode.
Instead, the sensors are designed with multiple input and output antennas and utilise the phase shifts over the receive antennas of incoming waveforms to estimate directions of arrival.
Furthermore, multiple radar sensors can be connected in a network to coherently fuse the radar data, thereby gaining a larger virtual aperture and an improved angle resolution.

The output of automotive RADAR sensors is a target list, comparable to a point cloud from a LiDAR, but commonly sparse in terms of point density.
However, RADAR systems also measure the instantaneous velocity of all detected targets, which can also power measures included as point information.




\section{Experiment}
\subsection{Experimental Platform}
\begin{figure}[hbt]
	\centering
	\begin{subfigure}{0.35\textwidth}
		\centering
		\includegraphics[width=0.9\linewidth]{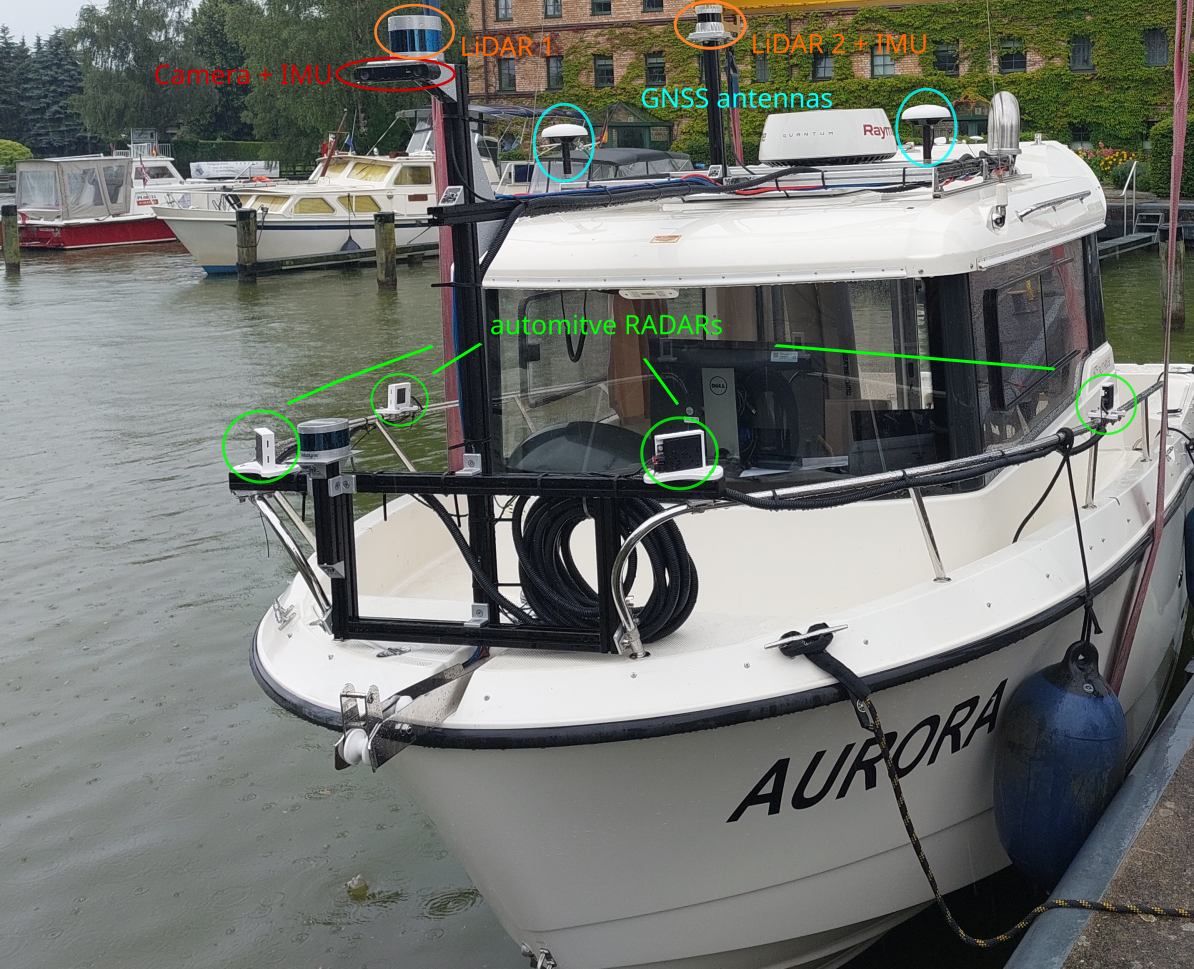}
		\caption{}
		\label{fig:aurora}
	\end{subfigure}
	\begin{subfigure}{0.35\textwidth}
		\centering
		\includegraphics[width=0.9\linewidth]{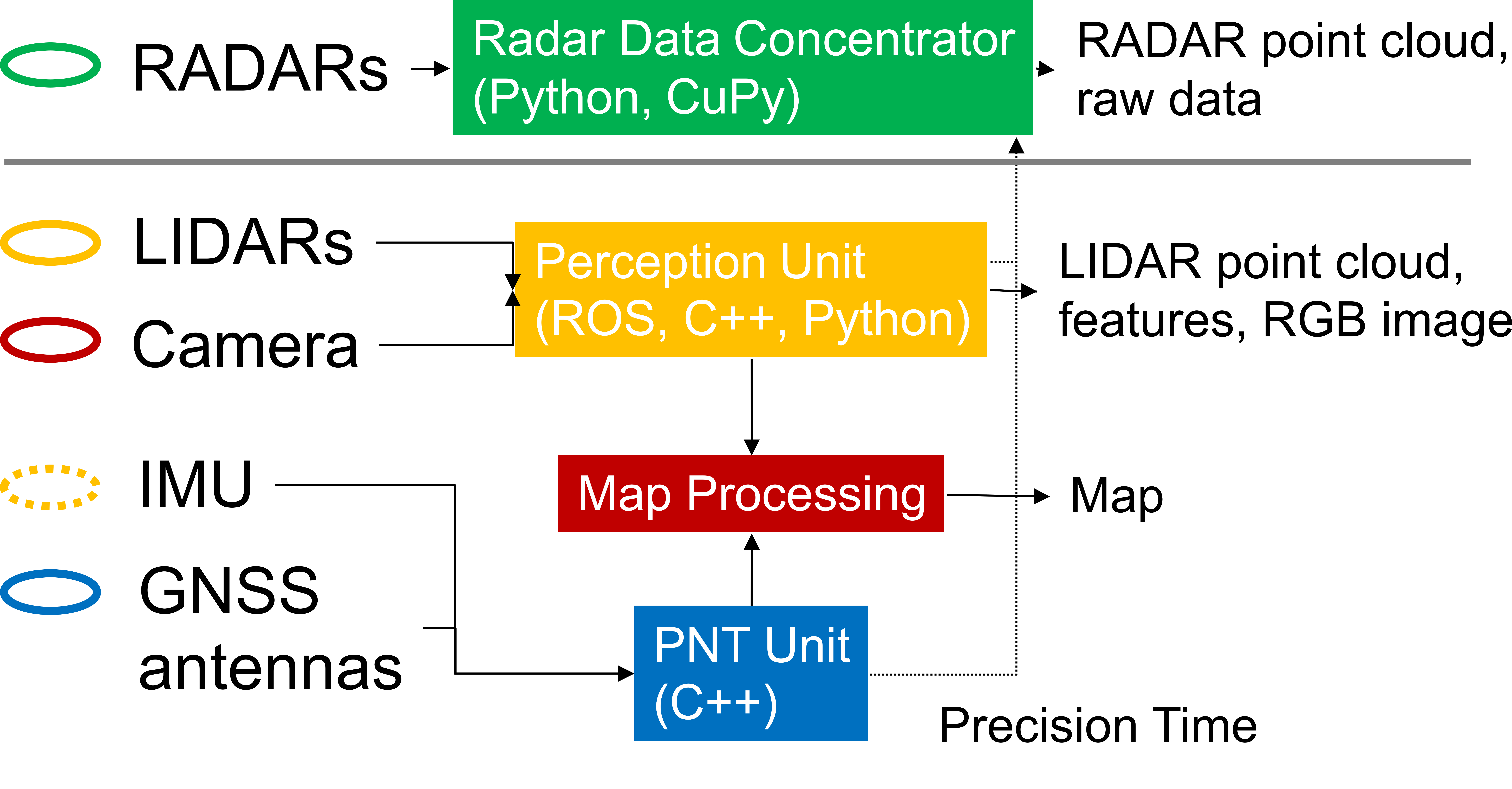}
		\caption{}
		\label{fig:aurora_block}
	\end{subfigure}
	\caption{A Fotograph of the experimental platform, the cabin boat Aurora, equipped with several sensors for navigation and enviromental perception (a) and
	  an illustrated blockdiagram showing an overview of the sensor inputs and their processing (b).}
		\label{fig:aurora_imgs}
\end{figure}

The cabin boat Aurora provided by the DLR, serves as an experimental platform.
Fig. \ref{fig:aurora} shows the boat equipped with several sensors.
The boat is equipped with two Global Navigation Satellite System (GNSS) antennas and inertial measurement units (IMU) integrated in a stereo camera and a LiDAR for precise localisation.
The camera captures semantic data, and three LiDARs record detailed geometric information about the vessel's surroundings.
Furthermore, a network of four automotive RADAR sensors are mounted at the front and sides of the boat, which also perceives the near-range environment of the boat.

As illustrated in the block diagram in Fig. \ref{fig:aurora_block}, the LiDAR sensors and the camera data are fed into the perception unit.
The Position, Navigation and Timing (PNT) unit receives geospatial information from the GNSS antennas and IMUs.
The LiDAR point cloud and camera image, along with the estimated vessel position, are used as input for the map processing unit.
The map processing unit's capabilities are fully detailed in \cite{hoesch2023}. 
It will be employed in future work for mapping and localisation using automotive RADAR sensor networks.
The PNT unit's precise GPS time is used to obtain the same time reference for LiDAR and RADAR data, which is essential for comparing both sensor outputs.
Fig. \ref{fig:aurora_block} also shows that the automotive RADARs are connected to a RADAR data concentrator (RDK).
The raw RADAR data is processed into 3D point clouds using the Python programming language.
We use the Python package CuPy to process and visualize the point clouds in real-time, taking full advantage of the RDK's graphics processing unit (GPU).
To develope and test various algorithms for RADAR data processing, the raw data is stored in packet capture (PCAP) files.

Tab. \ref{tab:sensorConfig} provides an overview of the sensor characteristics of the employed lidars and automotive radars.
There is a clear discrepancy between the resolutions of the lidars and the radar. 
While the lidars have an angle resolution well below one degree, the radar's resolution is much lower.
The low resolution is a directly result of the typically small aperture. 	
However, automotive radars are excellent at distinguishing targets based on their speed and, to a certain extent, their range.
This eases the need for good angle resolution, since a single target peak in the spatial domain is expected to be sufficient for detection.
Localising a global peak in the spatial domain can be achieved with high accuracy using peak interpolation. 
This can lead to angle estimates in ranges around one degree.
Nevertheless, super-resolution algorithms like MUSIC can achieve better angular resolution, 
but they demand more computing power. This may contradict the need for real-time performance.

\begin{table}[tbh]
	\centering
	\caption[LiDAR and RADAR specification parameters]{LiDAR and RADAR specification parameters}
	\label{tab:sensorConfig}
	\begin{threeparttable}
		\begin{tabular}{l c c c c}
			\toprule
			\multicolumn{2}{l}{Sensor unit} & LiDAR 1 & LiDAR 2 & RADAR \\
			\midrule
			Parameter & Unit & \multicolumn{3}{c}{Value} \\
			\midrule
			Range & $\mathrm{m}$ & $ \leq 200$ & $ \leq 75$ & $10 \leq 200$\\
			Range Resolution & $\mathrm{m}$ & $\leq 0.03$ & $\leq 0.01$ & $ 0.08 \leq 1.56$\tnote{1, 3}\\
			Azimuth field of View & ${}^{\circ}$ & $360$ & $360$ & $\approx 150$\\
			Elevation field of view & ${}^{\circ}$ & $-25$ to $15$ &  $\pm 45$& $\approx \pm 15$\\
			Azimuth Resolution & ${}^{\circ}$ & $0.1 \leq 0.4$ & $0.2 \leq 0.7$ & $\approx 6$\tnote{2, 3}\\
			Elevation Resolution & ${}^{\circ}$ & $0.1 \leq 0.4$ & $0.7 \leq 2.8$ & $\approx 35$\tnote{2, 3}\\
			Velocity & $\mathrm{m}\cdot\mathrm{s}^{-1}$ & & & $\leq \pm 28$\\
			Velocity & $\mathrm{m}\cdot\mathrm{s}^{-1}$ & & & $\leq \pm 0.1$\\
			Update rate & $\mathrm{Hz}$ & $5 \leq 20$ & $10$, $20$ & $\leq 20$\\
			\bottomrule
		\end{tabular}
		\begin{tablenotes}
			\item[1] Based on set chirp bandwidth and samples.
			\item[2] Receive array beam width.
			\item[3] Can be greatly improved by software
		\end{tablenotes}
	\end{threeparttable}
\end{table}

\subsection{FMCW RADAR Data Processing}
\begin{figure}[hbt]
	\centering
	\includegraphics[width=0.7\textwidth]{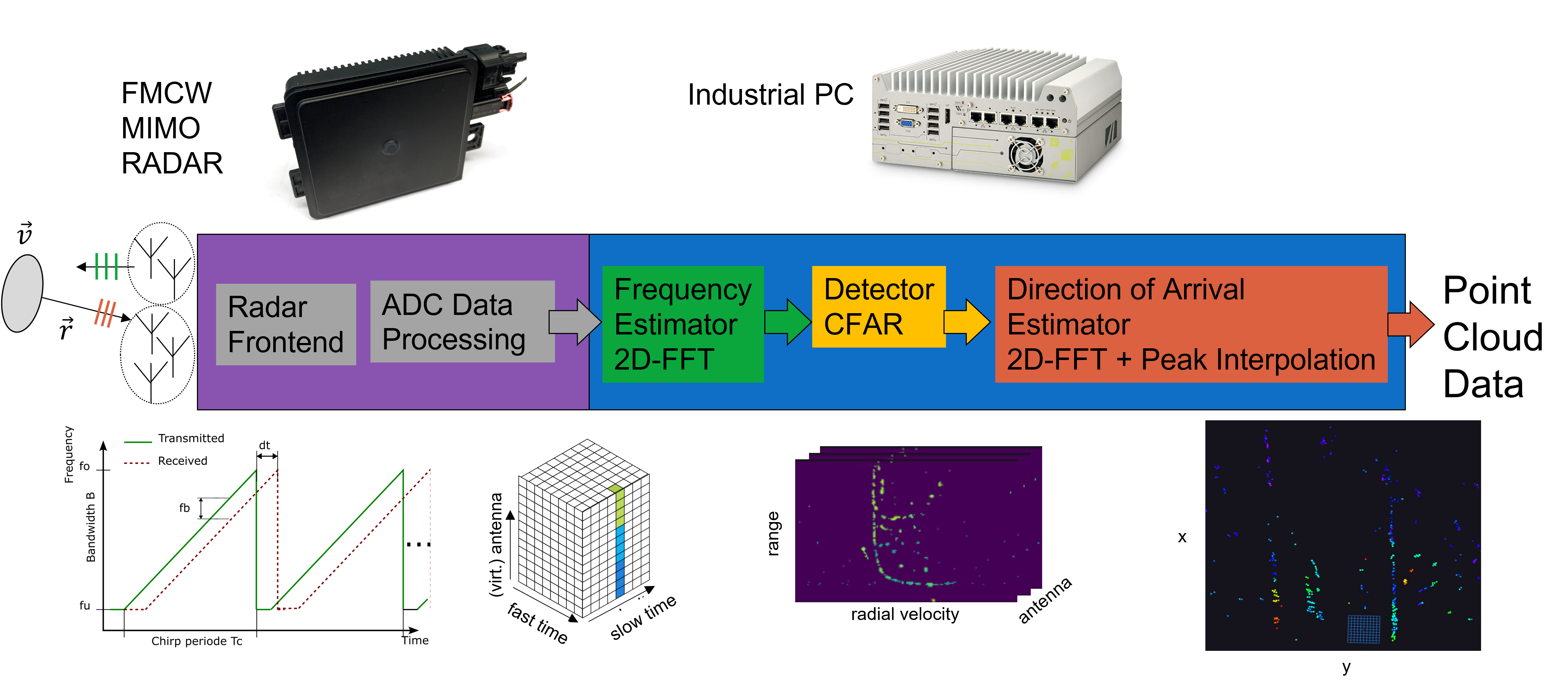}
	\caption{Overview of the RADAR signal processing pipeline of a single sensor.
	The purple box represents one of the used automotive RADAR sensors. The blue box contains
	the processing of the RADAR data concentrator (RDK) within the industrial PC }
	  \label{fig:radarProcPipeline}
 \end{figure}

This section provides an overview of the data processing of frequency-modulated continuous wave (FMCW) automotive radars.
In Fig. \ref{fig:radarProcPipeline} a block diagram illustrates the processing pipeline from the RADAR sensor's front end to the result of the RADAR point cloud.
The radar performs the tasks within the purple box, while the radar data concentrator (RDK) performs its tasks within the blue box.
The two devices are connected to each other via a Gigabit Ethernet connection.

At the beginning the RADAR frontend transmits frequency-modulated electromagnetic waveforms simultanously via three transmitting antennas utilising Doppler division multiple access (DDMA).
The waveforms consist of several short pulses swept over frequency, called chirps. 
They are illustrated in the bottom left of Fig. \ref{fig:radarProcPipeline}.

When a transmitted signal is scattered back from an object at position $\vec{r}$,
the reflected waveform is received by four antennas with a time delay, which corresponding to the object's radial range $|\vec{r}|$.
The RADAR front end mixes the received signal with the transmitted one to obtain an intermediate signal with a difference frequency proportional to the time delay.
Subsequently the analogue intermediate signal of each chirp is converted into the digital domain and fed to the radar data concentrator (RDK), which receives a data cube with the dimensions
fast-time, slow-time and antenna.
Fast-time describes the sampling within a single chirp signal, while slow-time defines the time intervals per chirp. 
The antenna dimension is a direct results of the multiple receive antennas.

To extract the mentioned difference frequency of the intermediate signal a Fourier transform is performed in fast-time domain.
Another Fourier transform along the slow-time axis extracts the Doppler frequency, 
which is a measure of the object's relative radial velocity, represented by the vector $\vec{v_r}$.
The Doppler frequency is the result of a periodic phase shift of the intermediate signal over all chirps for a fixed temporal position in fast-time.
Since noise is always added to our signals, we must use a detector to separate the object signals from the noise.

Once the 2D Fourier transform and detection step are complete, 
we can create maps that pinpoint the detected targets in the new dimensions range and velocity for each antenna, 
including virtual antennas.
The virtual antennas are the result of the three orthogonal transmission signals resulting from the DDMA approach.
Therefore, the four receive antennas receive signals from three independent transmitters, resulting in 12 virtual receive antennas.
Finally, we perform another Fourier transform for each detected target over the antenna dimension, which yields the angular information we need to localise a target.
Since we have estimated the range and angles of a target we can construct a point cloud, while each point in space holds information about signal strength and relative radial velocity.

\subsection{Measurement Campaigns}
\begin{figure}[hbt]
	\centering
	\begin{subfigure}{0.35\textwidth}
		\centering
		\includegraphics[width=0.9\linewidth]{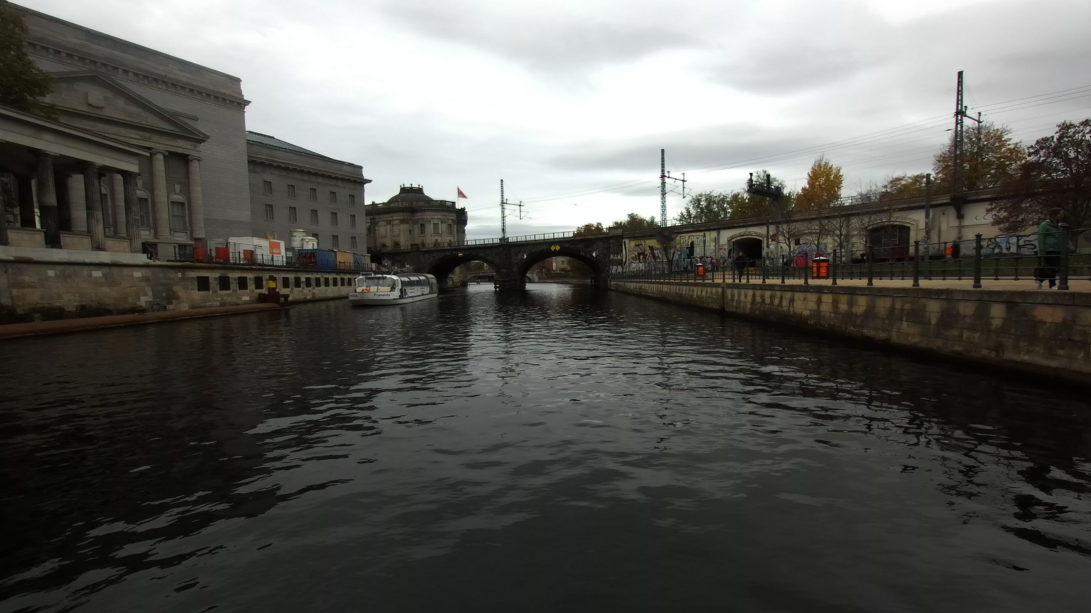}
		\caption{}
	  	\label{fig:sceneBerlin_cam}
	\end{subfigure}
	\begin{subfigure}{0.35\textwidth}
		\centering
		\includegraphics[width=0.9\textwidth]{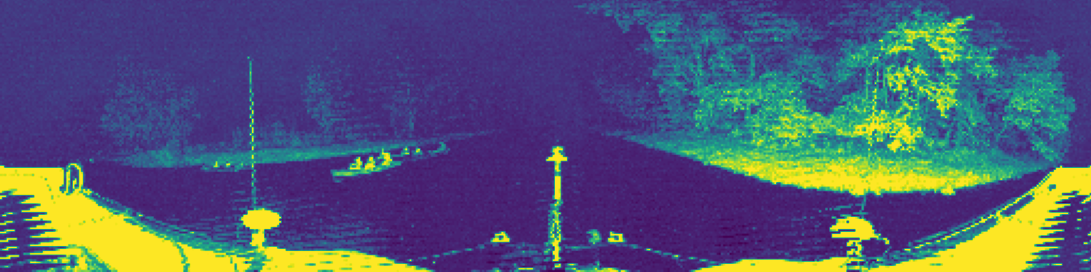}
		\caption{}
		\label{fig:sceneNeustrelitz_cam}
	\end{subfigure}
	\vfill
	\begin{subfigure}{.35\textwidth}
		\centering
		\includegraphics[width=0.9\linewidth]{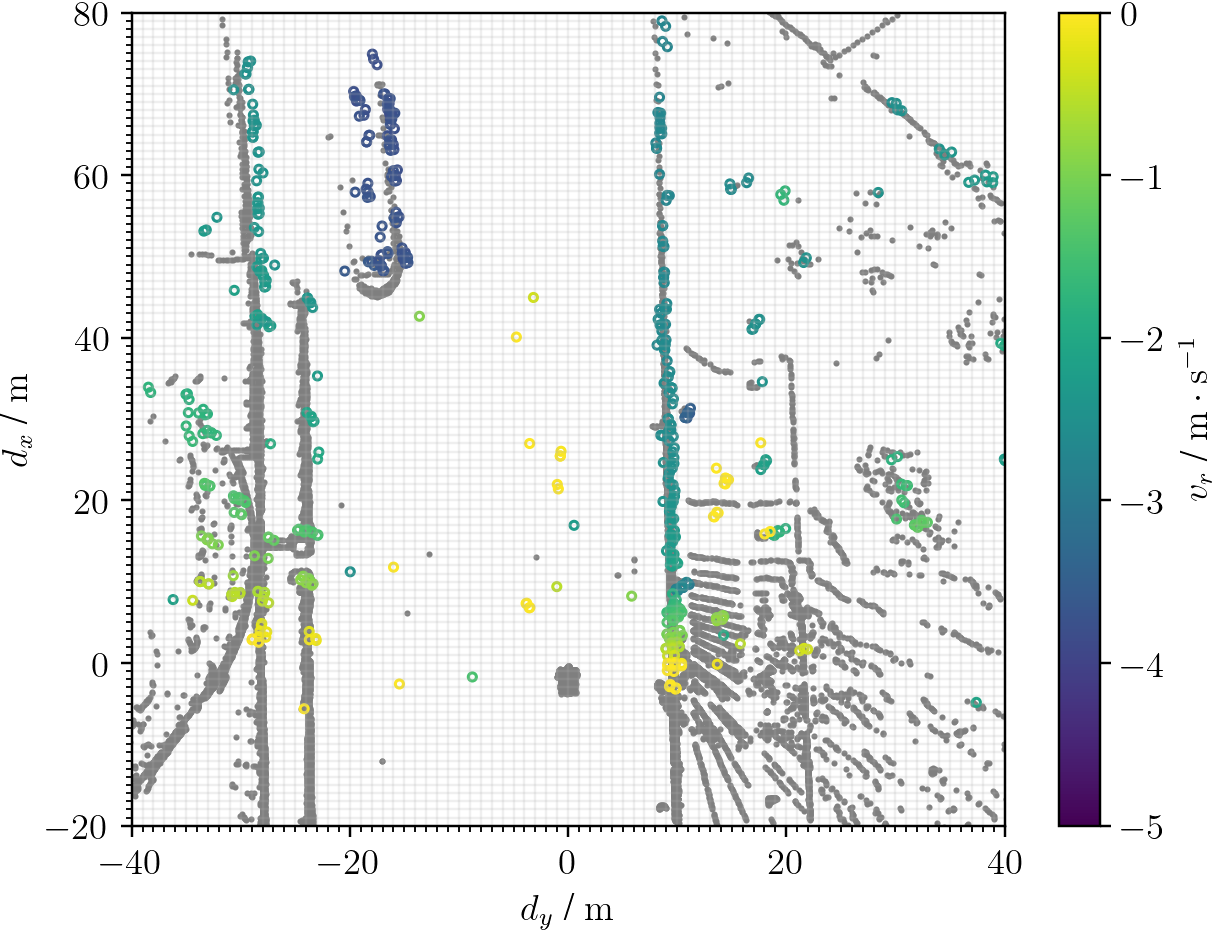}
		\caption{}
	  	\label{fig:sceneBerlin_LiDARRadar}
	\end{subfigure}
	\begin{subfigure}{.35\textwidth}
		\centering
		\includegraphics[width=0.9\linewidth]{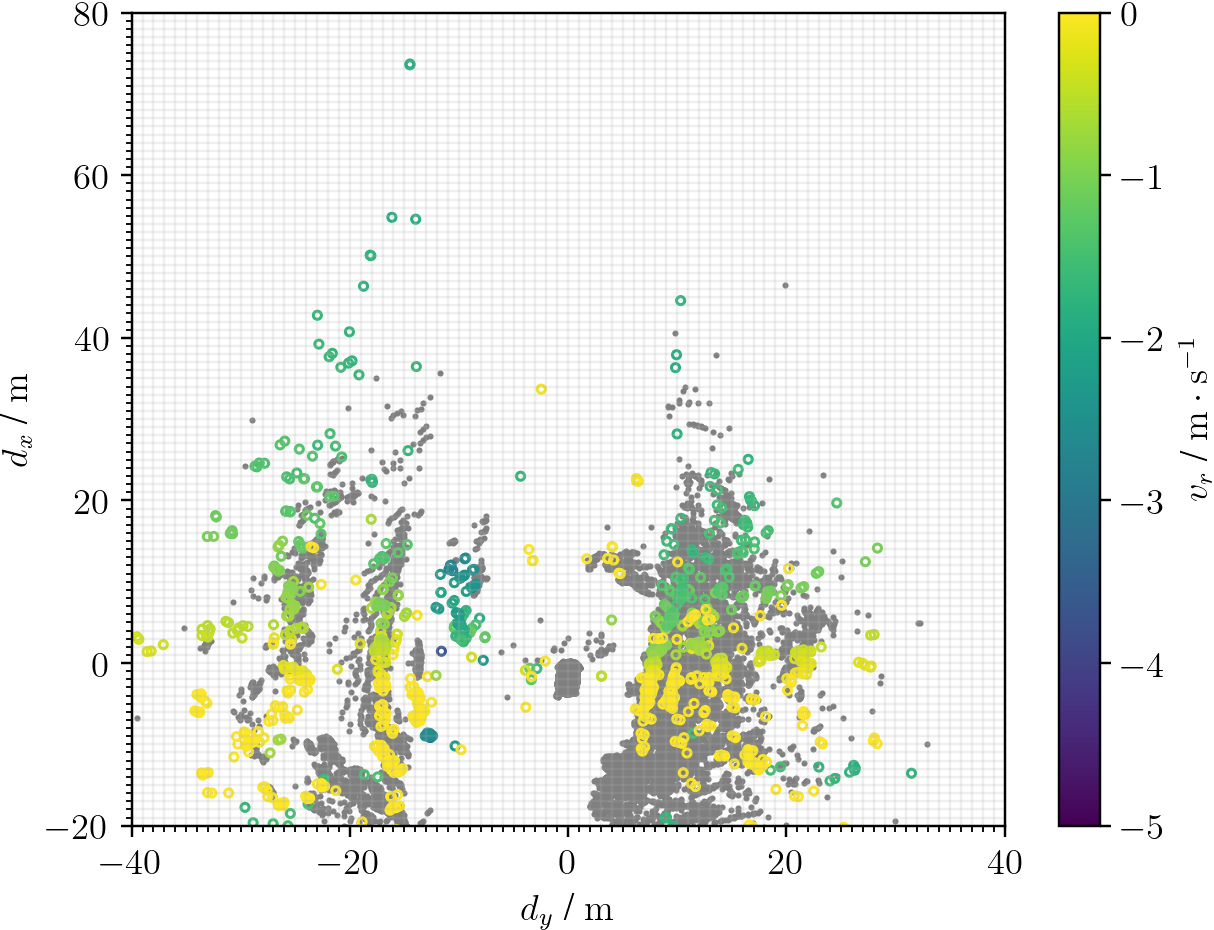}
		\caption{}
	  	\label{fig:sceneNeustrelitz_LiDARRadar}
	\end{subfigure}
	\caption{(a) Camera image of a selected scene from a scenario of the measurement campaign in Berlin, Germany.
	(b) LiDAR $360\,{}^{\circ}$ signal image of a selected scene from a scenario of the maesurement campaign in Neustrelitz, Germany.
	The camera image data was lost, hence the LiDAR image is used to visualize the scene.
	(c, d) Superimposed point clouds of LiDAR and  RADAR sensor network in birds eye's view of the corresponding scene with LiDAR points in gray
	and colored RADAR points indicating the relative radial velocity $v_r$}
	\label{fig:scenes}
 \end{figure}

TWe used two measurement campaigns carried out in Berlin and Neustrelitz, Germany, 
to compare the performance of LiDAR and RADAR. Each dataset was selected to provide a representative sample.
The experiments were conducted in two distinct inland waterway channel environments.
One took place in a feature-rich city environment surrounded by a lot of concrete infrastructure and 
some metal objects such as railings, as depicted in Fig. \ref{fig:sceneBerlin_cam}.
The other one was conducted in a natural setting with trees, 
bushes and grass along the waterway channel, as shown in Fig. \ref{fig:sceneNeustrelitz_cam}.
Fig. \ref{fig:sceneBerlin_LiDARRadar} and Fig. \ref{fig:sceneNeustrelitz_LiDARRadar} show the superposition of LiDAR and RADAR point clouds in top view for both scenes.

%



It must be acknowledged that the sensor setups employed in the two measurement campaigns were not identical.
In the Berlin campaign, a RADAR sensor network with two frontal sensors was used 
In contrast, in the Neustrelitz campaign four automotive radars were used.
Furthermore, different lidars have been used. With reference to \ref{tab:sensorConfig}, 
LiDAR 1 was used in Berlin and LiDAR 2 in Neustrelitz.

\subsection{Comparitive figures}
To provide a quantitative number for the point cloud densities, 
the scenes are transformed into 2D space, cropped to an area of 100 by 80 metres and 
divided into $M$ grid cells of one square metre in size.
Each point of the respective point cloud falling into a cell $m$ is counted separately for both LiDAR and RADAR.
The overall point count per cell $N_m$ is then averaged over all cells where $N_m$ is greater than one. 
This gives us the mean cell count $\overline{N}_{1\mathrm{m}^2}$ per square metre as given in \eqref{eq:count}.

\begin{align}
	\overline{N}_{1\mathrm{m}^2} = \frac{1}{M-K} \sum_{m=0}^{M} N_m
	\label{eq:count}
\end{align}

where $K$ is the number of cells where $N_m$ equals zero, either for LiDAR or RADAR. 

To give a quantitative figure of the matching between LiDAR and radar point clouds, the Jaccard coefficient $J$ is used.
This figure relates the sum of the cells where both the LiDAR $A$ and the RADAR sensor network $B$ detected targets to the overall sum of cells with detected targets. 
The relation is given in \eqref{eq:jaccard}. A value of $0$ indicates no similarity, while a value of $1$ represents a perfect match.

\begin{align}
	J(A,B) =\frac{| A \cap B |}{| A \cup B |}\, , \qquad 0 \leq J \leq 1
	\label{eq:jaccard}
\end{align}

%
%

\section{Results}

\begin{figure}[hbt]
	\centering
	\begin{subfigure}{.35\textwidth}
		\centering
		\includegraphics[width=0.9\linewidth]{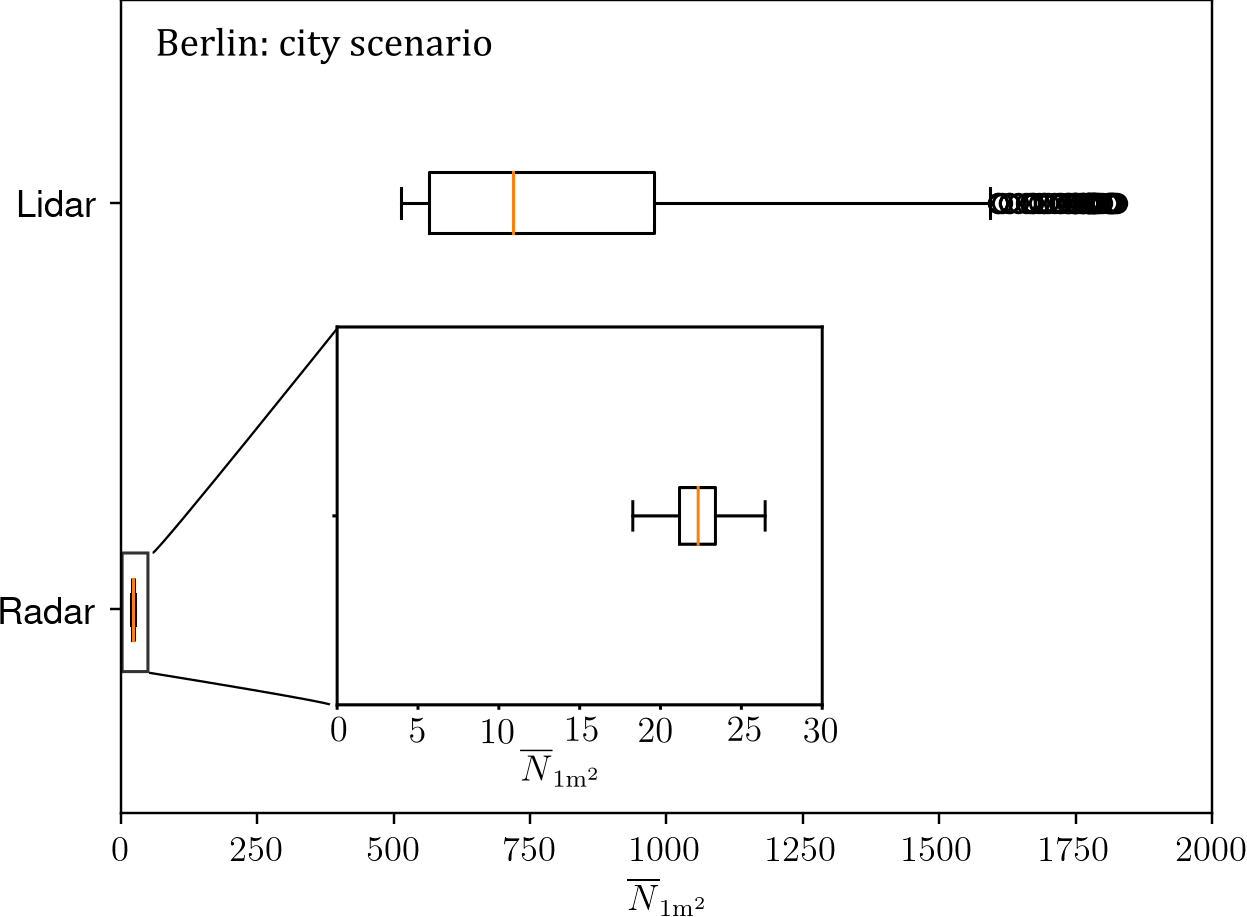}
		\caption{}
	  	\label{fig:sceneBerlin_count}
	\end{subfigure}
	\begin{subfigure}{.35\textwidth}
		\centering
		\includegraphics[width=0.9\linewidth]{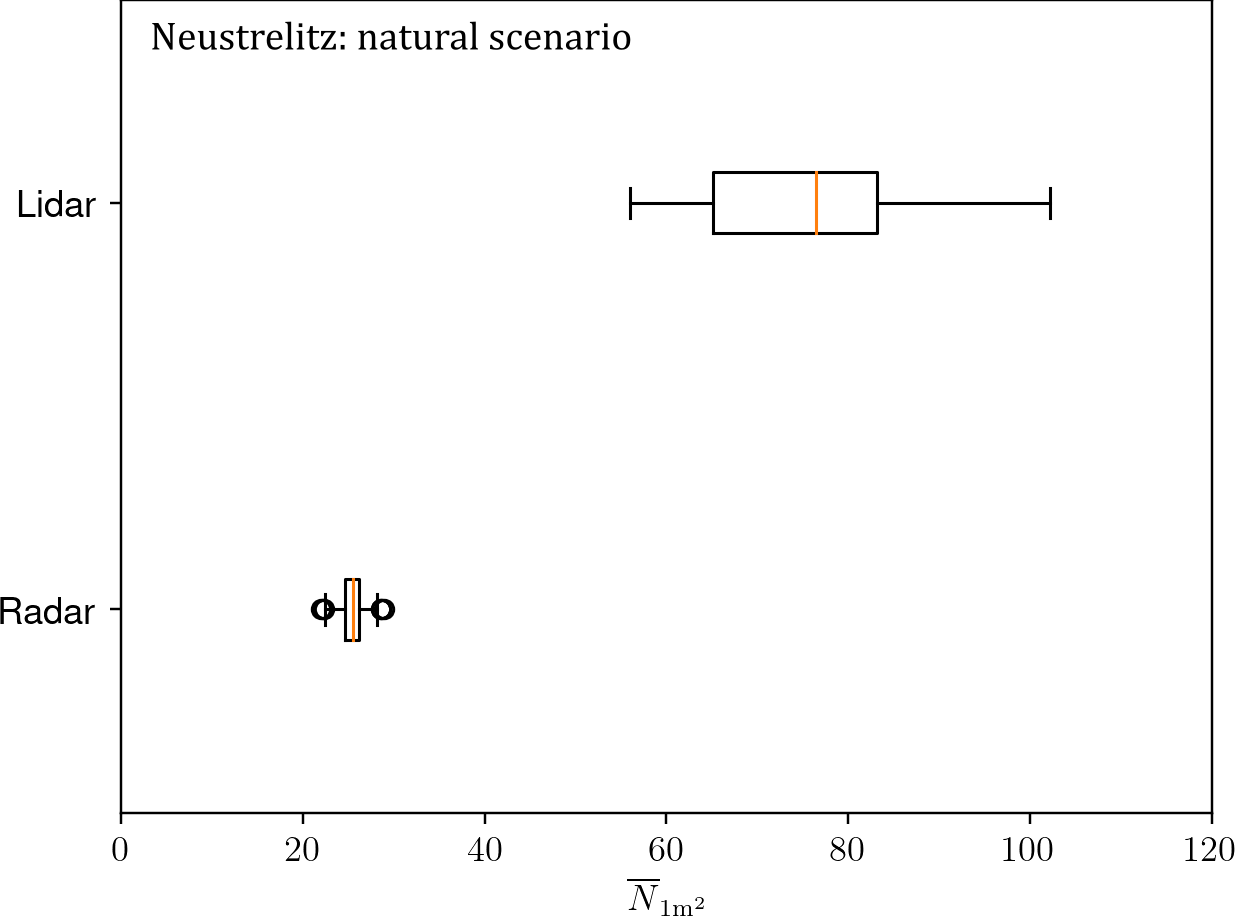}
		\caption{}
	  	\label{fig:sceneNeustrelitz_count}
	\end{subfigure}
	\caption{The distributions of the mean cell count $\overline{N}_{1\mathrm{m}^2}$ defined in \eqref{eq:count}
	 of the selected scenarios of the corrensponding measurement campaigns recorded in Berlin (a) and Neutrelitz (b). }
	\label{fig:count}
 \end{figure}

In this section the experimental results are quantitatively discussed.
First, the average cell counts defined in \eqref{eq:count} are presented in Fig. \ref{fig:count}.
The cell counts are collected over time for each secenario and summerized into a box plot.
For the city scenario obtained in Berlin the average cell count spreads evenly between about 15 and 30.
Compared to the corresponding LiDAR data this is a factor between 30 and 50 less.
This clearly shows the sparse density of radar point clouds.
In contrast to feature rich city enviroment, the difference between lidar and radar 
is much smaller in the more natural enviroment.
This is represented in Fig. \ref{fig:sceneNeustrelitz_count}. 
Here the ratio of collected targets per square meter cell between radar and lidar is about 3.
The significantly lower target number of the lidar could be caused by the vegetation, 
which prevents the narrow light beams passing between leaves from being reflected.
Looking at the distribution of the number of targets detected by the lidar, 
the Neutrelitz scenario is more centralized, 
which could also be a result of the rather random distribution of the vegetation structure.
The target detection of the automotive radar sensors 
appears to be more independed of the secenarios selected here.

\begin{figure}[hbt]
	\centering
	\begin{subfigure}{.35\textwidth}
		\centering
		\includegraphics[width=0.9\linewidth]{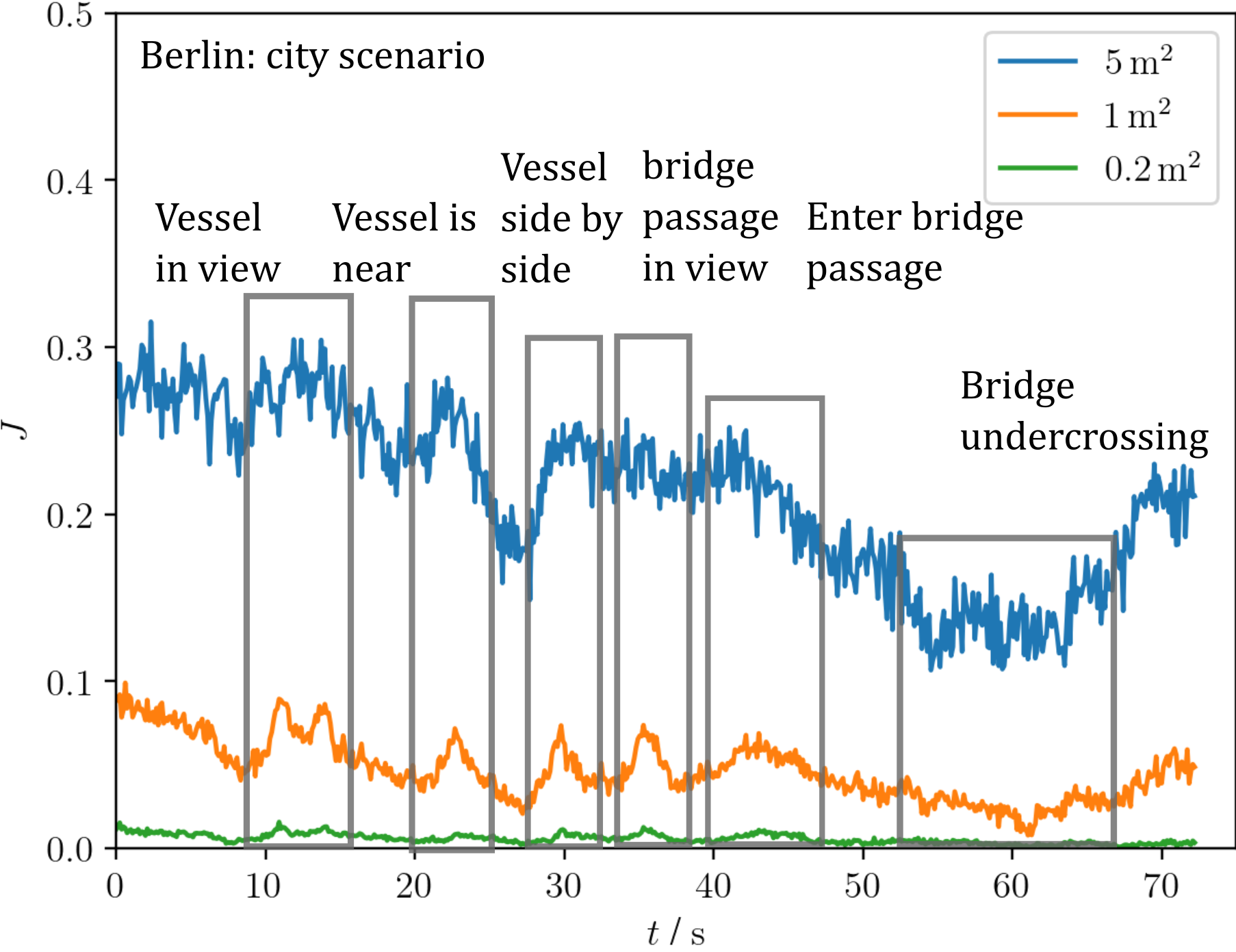}
		\caption{}
	  	\label{fig:sceneBerlin_jaccard}
	\end{subfigure}
	\begin{subfigure}{.35\textwidth}
		\centering
		\includegraphics[width=0.9\linewidth]{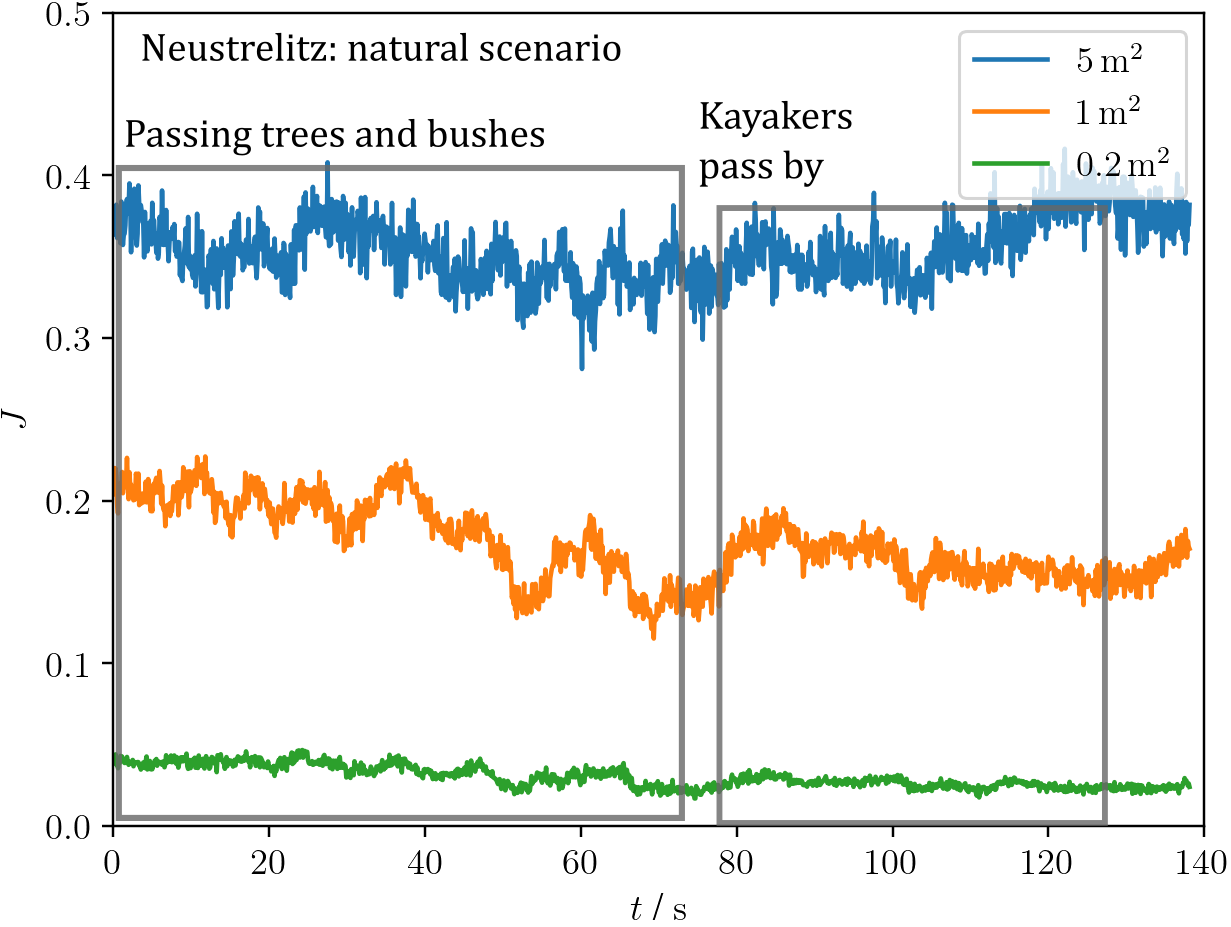}
		\caption{}
	  	\label{fig:sceneNeustrelitz_jaccard}
	\end{subfigure}
	\caption{Plotted Jaccard coefficient $J$ defined in \eqref{eq:jaccard} with
	 parameparameterized cell size over time 
	 of the corrensponding secenarios from Berlin (a) and Neutrelitz (b).
	 The scenarios are described by the gray rectangular boxes.}
	\label{fig:jaccard}
 \end{figure}

Now that the density of the point clouds has been compared, their similarity is examined.
In Fig. \ref{fig:sceneBerlin_jaccard} and Fig. \ref{fig:sceneNeustrelitz_jaccard} the Jaccarr similarity $J$ defined in \eqref{eq:jaccard}
 is plotted over time for the mentioned scenarios. 
Also certain events that take place during the recordings are highlighted.
Take into account Fig. , it is shown that the occurence of highly relecting targets in means of radar, 
like a vessel or guard rails in front of a bridge undercrossing, causing peaks of similarity.
Looking at Fig. \ref{fig:sceneBerlin_jaccard}, 
it can be seen that the occurrence of highly reflective targets in relation to radar, 
such as a ship or guard railss in front of a bridge undercrossings, leads to similarity peaks.
In general similarity between radar and lidar point clouds can be seen, but decreases with reduced cell size.
This can be caused by several reasons.
First of all, the error in the relative positions to the reference frame and time delays between the sensor recordings.
In addition, the lower accuracy of the automotive radar sensors leads to localization errors due to the compact sensor design, 
but above all due to the rather basic data processing used to date.
Furthermore, multipath effects can cause errors in radar point clouds.

\section{Conclusion and Outlook}

This paper presented a setup of a autmotive radar sensor network on a boat in combination with lidar.
To evaluate the autmotive RADAR sensors in inland waterway enviroments initial field tests were carried in two different scenarios.
The radar raw data was processed to point clouds, which were compared to point clouds measured by LiDAR.
As a quantitative comparative figures, a surface element based average target count and the Jaccard coefficient were used, 
which represent the density and similarity of lidar and radar point clouds.
Despite the lower radar point cloud density, the tested radar sensor network seems promising to perceive the surroundings of inland waterways.

For the future, time synchronization must be improved and algorithms with better accuracy for target localization must be investigated.
Furthermore, a method must be found to measure the positions of the radar sensors on board more precisely.
In the near future, we will investigate localization and mapping algorithms suitable for radar point clouds.

Finally, we want to realize an optimal modulation of the radar waveforms adapted to the surrounding situation, which leads to the topic of cognitive RADAR.
\section{Acknowledgment}
This work was supported by DLR, which primarily provided the Aurora cabin cruiser as a sensor platform and its sensor data.
The InnoSenT GmbH provided prototypes of one of their industrial automotive radar sensors and supported in the field of radar data processing.
This work was funded by the Federal Ministry for Digital and Transport as part of the DTW funding program.

\vsp{1em}


\begin{thebibliography}{88}
\def\bibname{LIT}
\bibitem{hoesch2023}
	L. Hösch, Y. Wellknown, A. Llorente, X. An, J. P. Llerena, and D. Medinaand, ``High definition mapping for inlandwaterways: Techniques, challenges and prospects'', \emph{2023 IEEE 26th International Conference on Intelligent Transportation Systems (ITSC)}, Bilbao, Spain, 2023 pp. 6034-6041
\end{thebibliography}

\vsp{1em}

\def\bibname{LIT}

\end{document}